# Exploring the Technical Knowledge Interaction of Global Digital Humanities: Three-decade Evidence from Bibliometric-based perspectives

Jiayi Li (2022201079@ruc.edu.cn), Renmin University of China, China, People's Republic of
Chengxi Yan (20218113@ruc.edu.cn), Renmin University of China, China, People's Republic of
Yurong Zeng (2022201115@ruc.edu.cn), Renmin University of China, China, People's Republic of
Zhichao Fang (fzc0225@163.com), Renmin University of China, China, People's Republic of
Huiru Wang (abcwhr2016@163.com), Renmin University of China, China, People's Republic of

## Introduction

**Digital Humanities (DH)** is an interdisciplinary field that integrates computational methods with humanities scholarship to investigate innovative topics. Each academic discipline follows a unique developmental path shaped by the topics researchers investigate and the methods they employ. With the help of bibliometric analysis, most of previous studies have examined DH across multiple dimensions such as research hotspots, co-author networks, and institutional rankings. However, these studies have often been limited in their ability to provide deep insights into the current state of technological advancements and topic development in DH. As a result, their conclusions tend to remain superficial or lack interpretability in understanding how methods and topics interrelate in the field.

To address this gap, this study introduced a new concept of **Topic-Method Composition (TMC)**, which refers to a hybrid knowledge structure generated by the co-occurrence of specific research topics and the corresponding method. Especially by analyzing the interaction between TMCs, we can see more clearly the intersection and integration of digital technology and humanistic subjects in DH. Moreover, this study developed a TMC-based workflow combining bibliometric analysis, topic modeling, and network analysis to analyze the development characteristics and patterns of research disciplines. By applying this workflow to large-scale bibliometric data, it enables a detailed view of the knowledge structures, providing a tool adaptable to other fields.

## Research Methodology

The methodology comprised three sequential analytical phases. First, we used topic keyword-based search method to collect DH-related documents from three different bibliometric databases (Web of Science, Crossref, and Dimension), applying rigorous data cleaning procedures to create a consolidated research corpus. The publication year was set between 1992 and 2022, which yielded a final dataset of 4,381 qualified articles for analysis.

Second, we developed specialized recognition models and optimization algorithms to accurately extract knowledge elements from the corpus. For method elements, we used the two-stage recognition algorithm (Li / Yan 2023) denoted as LLMRule-based. To be specific, a large language model (LLM) like ChatGPT performed preliminary method entity extraction, followed by rule-based standardization using an evolving domain dictionary. Let $i$ represent a method element, we calculated the number of documents that has used such method (i.e. $D_i$). And for topic elements, we used the state-of-the-art model Bertopic (Grootendorst 2022), which utilized contextual embeddings and clustering techniques to uncover nuanced topic structures. Model performance was optimized through cluster numbers, and two experts were invited to validate high-scoring terms and ensure interpretable topic labels. Hence for a certain topic $j$, we can obtain the total number $D_j$ of documents it has, as the formula (1).

Finally, we were able to automatically generate TMCs and construct a network of these compositions. The TMCs were represented as a global topic-method bipartite graph, where co-occurrence relationships mapped the associations between topics and methods. The normalized edge weights reflected to what extent this method was utilized in the given topic. Assuming $D_{i,j}$ as number of co-occurring documents for a method $i$ and topic $j$, $C_{i,j}$ as the interactive intensity between them, we used truncation strategies to preserve the pair link when $C_{i,j}$ is above a certain parameter $\sigma$. Through two to three rounds of threshold adjustments along with expert validation, optimal results were achieved at $\sigma = 0.001$. The relationship of TMC $R_{i,j}$ can be measured as the formula (2).

$$D_j = \left\{ d \middle| \underset{d}{\mathrm{argmax}}\, z(j|d) \right\} \quad (1)$$

$$R_{i,j} = \begin{cases} C_{i,j} & \text{if } C_{i,j} = \dfrac{D_{i,j}}{D_i \cdot D_j} > \sigma \\ 0 & \text{else} \end{cases} \quad (2)$$

Figure 1. Formulas used in this study

## Experiment

To evaluate our LLMRule-based algorithm, we randomly extracted 800 DH papers and compared two unsupervised approaches: LLM-based (using GPT-3.5-Turbo with instructional prompts) and Rule-based (employing an expert-curated dictionary with method synonyms and morphological variants). As shown in Table 1, LLMRule-based outperformed both baselines, demonstrating the superior integration of LLM capabilities and domain-specific method knowledge.

Table 1. Comparative evaluation for method recognition approaches

| Approach | Precision (%) | Recall (%) | F1 (%) |
|---|---|---|---|
| LLM-based | 20.62 | 11.98 | 15.15 |
| Rule-based | 18.28 | 33.95 | 23.77 |
| LLMRule-based | 43.69 | 26.14 | 32.71 |

Next, to ensure the interpretability and relevance of the generated topics for Bertopic, we employed perplexity (Kobayashi 2014) and coherence value (Zhang et al. 2022) metrics, which are widely used for the assessment of the quality of topic models. Ultimately, optimal performance occurred at 38 topics, where perplexity minimized and coherence maximized. And expert evaluation confirmed these topics were both statistically robust and conceptually meaningful.

Like typical bibliometric networks, the TMC network consisted of nodes (topics/methods) and edges representing shared elements between TMCs. When two TMCs shared a topic or method, they were connected. Then we employed two analytical approaches: (1) frequency analysis to identify prevalent TMCs, and (2) modularity-based community detection (Newman 2004) to reveal clustered knowledge interactions between topics and methods.

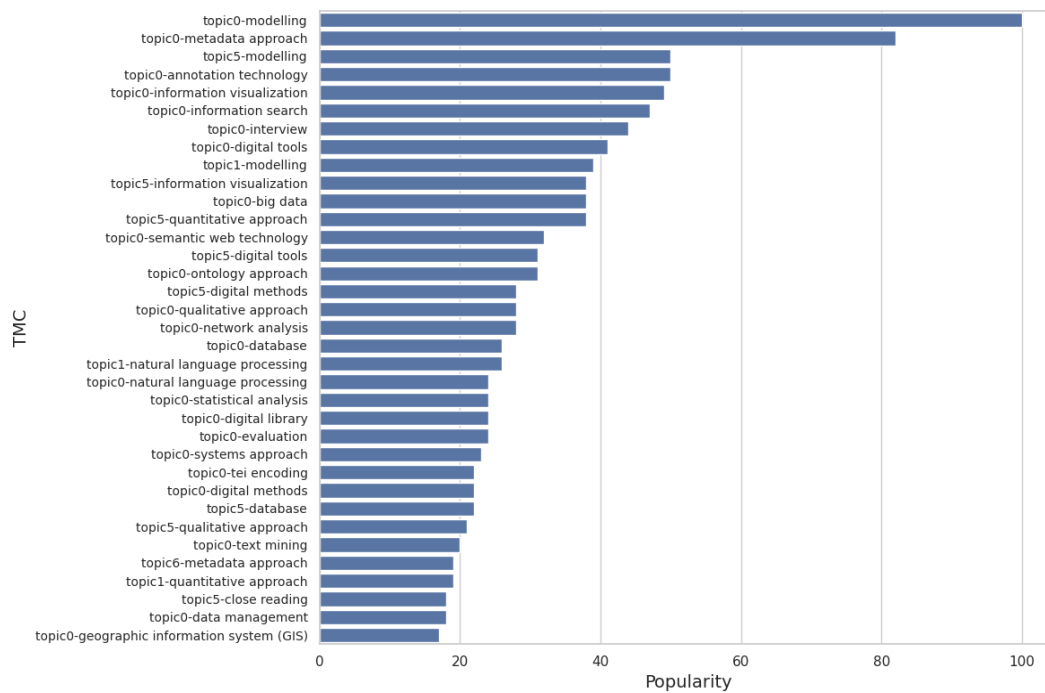

Figure 2. Top 35 TMCs by popularity

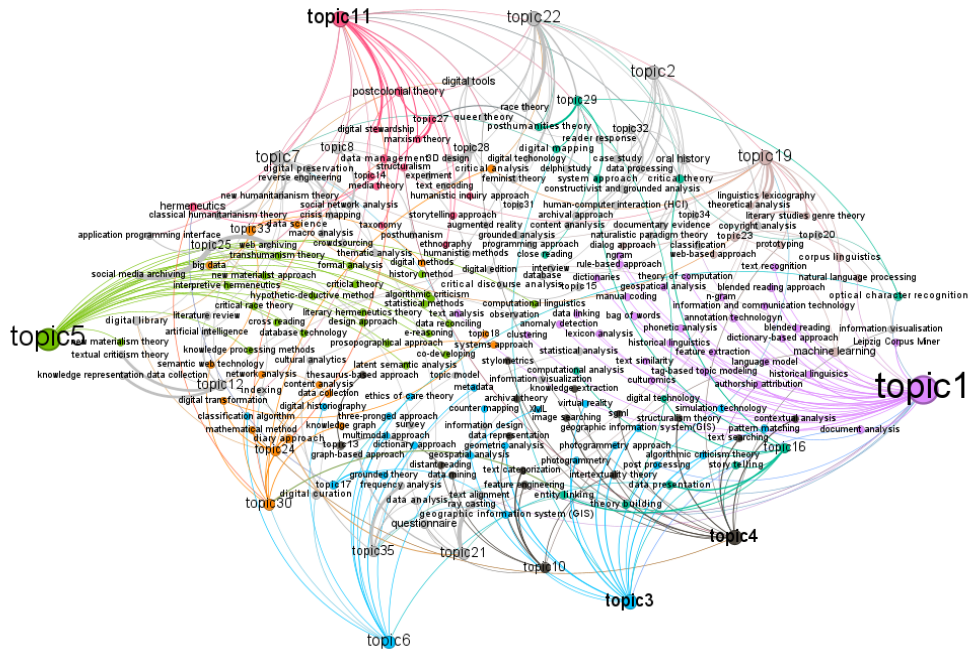

Figure 3. TMC network visualization

The findings underscore the interdisciplinary nature of DH, highlighting how computational methods—such as text mining, natural language processing, data visualization, and GIS—complement traditional humanities approaches like linguistic analysis, cultural heritage studies, and literary criticism. By analysing this TMC network, we identified 9 key research paradigms in DH through network clustering, revealing the diverse ways in which DH topics intersect with computational methods. These communities, derived from network clustering, highlight the diversity of research approaches within DH. For instance, some communities focus on linguistic and computational techniques (e.g., text mining, n-grams), while others emphasize the use of digital tools in cultural heritage studies or spatial analysis in archaeology. These communities reflect how DH research spans across a wide array of methodologies and topics, demonstrating the multidimensional and interdisciplinary nature of the field.

## Conclusion

The TMC framework provides a comprehensive approach to understanding the structure of DH research, offering valuable tools for identifying research gaps, improving topic-method alignment, and fostering collaboration within the field. Furthermore, the workflow is not limited to DH alone; it can be applied to other academic disciplines to explore the relationships between research topics and methods in a variety of fields. This makes the methodology highly versatile and applicable for a wide range of interdisciplinary studies.

## Acknowledgement

The study is supported and funded by the National Natural Science Foundation of China (No. 72204258). The corresponding author is Chengxi Yan (ORCID: 0000-0003-1128-550X, Email: 20218113@ruc.edu.cn).